\documentclass[conference]{IEEEtran}
\IEEEoverridecommandlockouts

\makeatletter
\newcommand{\linebreakand}{%
  \end{@IEEEauthorhalign}
  \hfill\mbox{}\par
  \mbox{}\hfill\begin{@IEEEauthorhalign}
}
\makeatother

\usepackage[export]{adjustbox}
\usepackage{booktabs}
\usepackage{float}
\usepackage{flushend}
\usepackage{graphicx}
\usepackage{ifpdf}
\usepackage[utf8]{inputenc}
\usepackage{listings}
\usepackage{multirow}
\usepackage{url}
\usepackage{tabularx}
\usepackage{textcomp}
\usepackage{tikz}
\usepackage{xcolor}
\usepackage{xspace}
\usepackage{comment}
\usepackage{subcaption}
\usepackage[htt]{hyphenat}
\usepackage{hyperref}
\usepackage[nameinlink]{cleveref}

\newif\ifdraft{}

\ifdraft{}
 \newcommand{\aanote}[1]{ {\textcolor{green} { ***aymen: #1 }}}
 \newcommand{\asnote}[1]{ {\textcolor{purple} { ***shao: #1 }}}
 \newcommand{\apnote}[1]{ {\textcolor{green} { ***andrew: #1 }}}
  \newcommand{\mtnote}[1]{ {\textcolor{orange} { ***matteo: #1 }}}
  \newcommand{\jhanote}[1]{ {\textcolor{red} { ***shantenu: #1 }}}
  \newcommand{\miknote}[1]{ {\textcolor{brown} { ***mikhail: #1 }}}
  \newcommand{\amnote}[1]{ {\textcolor{blue} { ***andre: #1 }}}
  \newcommand{\ooknote}[1]{ {\textcolor{blue} { ***ozgur: #1 }}}
  \newcommand{\twnote}[1]{ {\textcolor{blue} { ***tianle: #1 }}}
 \newcommand{\pjmnote}[1]{ {\textcolor{blue} { ***pete: #1 }}}
  \newcommand{\generalnote}[1]{ {\textcolor{gray} { *note: #1 }}}

\else
  \newcommand{\aanote}[1]{}
  \newcommand{\asnote}[1]{}
  \newcommand{\apnote}[1]{}
  \newcommand{\mtnote}[1]{}
  \newcommand{\jhanote}[1]{}
  \newcommand{\miknote}[1]{}
  \newcommand{\amnote}[1]{}
  \newcommand{\ooknote}[1]{}
  \newcommand{\twnote}[1]{}
  \newcommand{\generalnote}[1]{}
  \newcommand{\pjmnote}[1]{}
\fi

\ifpdf{}
  \DeclareGraphicsExtensions{.pdf, .jpg, .tif}
\else
  \DeclareGraphicsExtensions{.eps, .jpg, .ps}
\fi

\newlength\myheight
\newcommand*\circled[1]{\settowidth{\myheight}{#1}%
    \raisebox{-.1\myheight}{\tikz[baseline=(char.base)]{%
        \node[shape=circle,draw,minimum size=\myheight*\myheight*.4,inner sep=1pt](char){#1};}}}

\Crefname{figure}{Fig.}{Figs.}
\usepackage[normalem]{ulem}

\begin{document}

\title{RHAPSODY: Execution of Hybrid AI-HPC Workflows at Scale}
\iftrue
\author{\IEEEauthorblockN{Aymen Alsaadi}
 \IEEEauthorblockA{Rutgers University \\
 New Brunswick, NJ, USA \\
 0000-0001-7491-4946}
 \and
 \IEEEauthorblockN{Mason Hooten}
 \IEEEauthorblockA{Rutgers University \\
 New Brunswick, NJ, USA \\
 0000-0001-6321-8267}
 \and
 \IEEEauthorblockN{Mariya Goliyad}
 \IEEEauthorblockA{Rutgers University \\
 New Brunswick, NJ, USA \\
 0000-0002-5804-1935}
 \linebreakand
 \IEEEauthorblockN{Andre Merzky}
 \IEEEauthorblockA{Rutgers University \\
 New Brunswick, NJ, USA \\
 0000-0002-7228-4327}
 \and
 \IEEEauthorblockN{Andrew Shao}
 \IEEEauthorblockA{Hewlett Packard Enterprise Canada\\
 Victoria, BC, Canada \\
 0000-0003-3658-512X}
 \and
 \IEEEauthorblockN{Mikhail Titov}
 \IEEEauthorblockA{Brookhaven National Laboratory \\
 Upton, NY, USA \\
 0000-0003-2357-7382}
 \linebreakand
 \IEEEauthorblockN{Tianle Wang}
 \IEEEauthorblockA{Brookhaven National Laboratory \\
 Upton, NY, USA \\
 0000-0001-8293-0671}
 \and
 \IEEEauthorblockN{Yian Chen}
 \IEEEauthorblockA{Hewlett Packard Enterprise \\
 Palo Alto, CA, USA \\
 0009-0000-5074-6458}
 \and
 \IEEEauthorblockN{Maria Kalantzi}
 \IEEEauthorblockA{Hewlett Packard Enterprise \\
 Palo Alto, CA, USA \\
 0000-0003-0014-5480}
 \linebreakand
 \IEEEauthorblockN{Kent Lee}
 \IEEEauthorblockA{Hewlett Packard Enterprise \\
 Palo Alto, CA, USA \\
 0009-0001-2616-0817}
 \and
 \IEEEauthorblockN{Andrew Park}
 \IEEEauthorblockA{Rutgers University \\
 New Brunswick, NJ, USA \\
 0009-0000-3468-0049}
 \and
 \IEEEauthorblockN{Indira Pimpalkhare}
 \IEEEauthorblockA{Hewlett Packard Enterprise \\
 Palo Alto, CA, USA \\
 0000-0002-2950-5606}
 \linebreakand
 \IEEEauthorblockN{Nick Radcliffe}
 \IEEEauthorblockA{Hewlett Packard Enterprise \\
 Palo Alto, CA, USA \\
 0009-0005-4703-8959}
 \and
 \IEEEauthorblockN{Colin Wahl}
 \IEEEauthorblockA{Hewlett Packard Enterprise \\
 Palo Alto, CA, USA \\
 0000-0003-3585-3401}
 \and
 \IEEEauthorblockN{Pete Mendygral}
 \IEEEauthorblockA{Hewlett Packard Enterprise \\
 Palo Alto, CA, USA \\
 0009-0009-3643-5093}
 \linebreakand
 \IEEEauthorblockN{Matteo Turilli}
 \IEEEauthorblockA{Rutgers University \\
 New Brunswick, NJ, USA \\
 IE University \\
 Segovia, Segovia, Spain \\
 0000-0003-0527-1435}
 \and
 \IEEEauthorblockN{Shantenu Jha}
 \IEEEauthorblockA{Rutgers University \\
 New Brunswick, NJ, USA \\
 Princeton Plasma Physics Laboratory \\
 Princeton, NJ, USA \\
 0000-0002-5040-026X}
 }
\fi 

\maketitle

\begin{abstract}
Hybrid AI-HPC workflows combine large-scale simulation, training, high-throughput inference, and tightly coupled, agent-driven control within a single execution campaign. These workflows impose heterogeneous and often conflicting requirements on runtime systems, spanning MPI executables, persistent AI services, fine-grained tasks, and low-latency AI-HPC coupling. Existing systems typically address only subsets of these requirements, limiting their ability to support emerging AI-HPC applications at scale. We present RHAPSODY, a multi-runtime middleware that enables concurrent execution of heterogeneous AI-HPC workloads through uniform abstractions for tasks, services, resources, and execution policies. Rather than replacing existing runtimes, RHAPSODY composes and coordinates them, allowing simulation codes, inference services, and agentic workflows to coexist within a single job allocation on leadership-class HPC platforms. We evaluate RHAPSODY with Dragon and vLLM on multiple HPC systems using representative heterogeneous, inference-at-scale, and tightly coupled AI-HPC workflows. Our results show that RHAPSODY introduces minimal runtime overhead, sustains increasing heterogeneity at scale, achieves near-linear scaling for high-throughput inference workloads, and data- and control-efficient coupling between AI and HPC tasks in agentic workflows.
\end{abstract}


\section{Introduction}\label{sec:intro}

Hybrid AI-HPC workflows fuse the learning capabilities of AI with the physical accuracy of High-Performance Computing~\cite{brewer2024ai}. This synergy is crucial for modern discovery, as it drastically reduces time and energy required to solve problems---such as screening millions of potential drug compounds or designing new materials---that would be computationally infeasible using traditional methods alone~\cite{jha2023ai}. For example, instead of relying solely on slow, brute-force simulations, these workflows utilize AI models as "surrogates" to approximate complex calculations or to intelligently steer simulations toward the most promising experimental paths in real time. 

The coupling of AI and HPC tasks in hybrid AI-HPC workflows manifests in three primary modes: AI-in-HPC, which embeds surrogate models to accelerate core physics; AI-out-HPC, which enables real-time inference and digital twins; and AI-around-HPC, which uses AI to orchestrate and optimize complex campaigns. These approaches necessitate a fundamental shift in the design of underlying workflow systems and runtime technologies to support the complex interplay between simulation and learning tasks.

Concurrently supporting the triad of simulation, training, and inference is critical in many workflows, but particularly so in the development and application of scientific foundation models (FM). The importance of concurrent simulation, training, and inference lies in generating physically deterministic data in real time, thereby shifting the learning paradigm from training on static data to dynamic causal interactions. In this concurrent loop, the simulation not only provides a dataset but also actively reacts to the model’s inference steps, calculating immediate physical consequences. By coupling the model’s outputs with simulations in an ``active learning'' loop, the system produces a continuous stream of physically grounded, interactive experiences, enabling the FM to internalize the underlying governing equation. 

The American Science Cloud (AmSC)---an early prototype of the Genesis Mission~\cite{doe_genesis_website} Platform---is a unified, federated DOE platform that integrates computing facilities and experimental data resources to build and use scientific foundation models. It serves as a backbone that enables AI to run where scientific instruments and simulation data are generated, allowing researchers to perform scalable model training and inference across distributed sites. Thus, in addition to building FM for diverse scientific domains, the AmSC supports various workflows, such as prompt-driven inference loops that chain simulations and evaluations to iteratively produce scientific outcomes, and agentic orchestration, in which AI agents autonomously control the discovery and execution cycle. 


Effectively running these integrated workflows requires a runtime system capable of handling extreme heterogeneity, including long-running MPI-based simulation executables (e.g., GROMACS), high-throughput inference hosted on distributed systems such as vLLM, and tightly coupled AI–HPC tasks that require low-latency data exchange and coordinated resource orchestration~\cite{tummalapalli2025transit}. These workflows also rely on Model Context Protocol (MCP) servers to enable interoperability and autonomous, agent-driven management. As a result, the runtime requirements of hybrid AI–HPC workflows are non-trivial in both functionality and performance.

\jhanote{this  paragraph can be eliminated or significantly reduced}\mtnote{Done.}


This paper presents RHAPSODY, a multi-runtime workflow execution substrate that adopts a building-blocks approach: rather than replacing specialized systems, RHAPSODY composes them behind uniform abstractions for tasks, resources, and execution policies. RHAPSODY integrates runtime technologies for HPC task execution (e.g., Flux and Dragon) with AI-serving and training frameworks (e.g., vLLM and DeepSpeed), and treats long-running services and their client interactions as first-class elements of workflow execution. This design enables workflows to mix executables, service-based inference/training, and fine-grained AI-HPC interactions without forcing artificial phase boundaries or embedding backend-specific logic into workflow code.

We characterize and evaluate RHAPSODY using representative hybrid AI-HPC use cases spanning simulation, training, and inference. Our experiments quantify runtime overheads and characterize workflow performance in terms of heterogeneity, throughput, resource utilization, coupling, and scaling behavior on leadership-class HPC platforms. The results show that RHAPSODY sustains concurrent heterogeneous execution with low overhead while improving effective utilization across diverse workflow structures.

The contributions of this paper are fourfold: (1) We present the design of RHAPSODY, a runtime system for heterogeneous hybrid AI-HPC workflows; (2) we demonstrate how RHAPSODY composes existing runtime building blocks for executables, inference, and tightly interacting AI-HPC tasks; (3) we validate RHAPSODY through representative scientific use cases that demand heterogeneous task execution; and (4) we characterize RHAPSODY's performance, showing low runtime overheads and high efficiency at scale.

The remainder of the paper is organized as follows: Section~II introduces the motivating workflow classes and use cases used in our evaluation; Section~III describes the design and implementation of RHAPSODY; and Section~IV presents experimental results and performance characterization.

\section{Use Cases and Motivating Workflows}\label{sec:uc}

\mtnote{All: I rewrote the opening from scratch. Please edit/iterate as needed.}

The design of RHAPSODY is driven by scientific workflows that integrate traditional HPC and AI methods within a single execution campaign. \jhanote{I think the following arguments have been made well earlier. From ... }
\jhanote{... till here.}\mtnote{Agreed.} To ground RHAPSODY’s design in concrete application requirements, we identify three representative categories of hybrid AI–HPC workflows that capture dominant integration patterns observed in contemporary scientific endeavors. While real workflows often combine multiple categories, these exemplars, together, highlight the core runtime capabilities required to support heterogeneous execution, scalable inference, and tightly coupled AI–HPC interaction on leadership-class systems.

\subsection{Heterogeneous Workloads}
\label{ssec:uc-heterogeneity}


Heterogeneous scientific workflows integrate task types that differ substantially in computational structure, resource requirements, and execution dynamics. Typical workflows combine CPU-bound analytics, GPU-accelerated machine-learning kernels, large-scale MPI simulations, data preprocessing stages, and surrogate or scoring models. These tasks span multiple programming models and granularities, requiring a runtime system capable of dynamically scheduling mixed CPU/GPU workloads, sustaining high concurrency, and adapting to asynchronous task generation.

The \emph{Integrated Modeling PipEline for COVID Cure by Assessing Better LEads} (IMPECCABLE) campaign exemplifies this class. IMPECCABLE accelerates drug-discovery pipelines by combining physics-based simulations with AI-driven surrogate models and generative components. Its workflows include large batches of molecular dynamics simulations, ML-based docking surrogates for candidate ranking, scoring and refinement stages that mix CPU-dominated preprocessing with GPU-accelerated evaluation, and iterative active-learning loops that guide subsequent sampling. These tasks vary by orders of magnitude in runtime, memory, and resource modality, ranging from small, short-lived analysis tasks to large MPI simulations spanning wide ranges of ranks.

Beyond task diversity, IMPECCABLE workflows exhibit complex temporal structure: multiple pipeline stages execute concurrently, leading to situations in which simulations, ML models, data transformations, and filtering tasks occupy different subsets of resources simultaneously. Efficient execution therefore requires robust task placement across mixed resources, runtime adaptivity as new tasks are generated, and high throughput in the presence of many small tasks interleaved with large MPI jobs or GPU-intensive kernels.

For RHAPSODY, heterogeneous workloads motivate a unified runtime model that can manage task diversity without specialized orchestrators for individual stages. The runtime must preserve dependency structures across modalities, ensure efficient resource sharing, and maintain throughput despite wide variation in task granularity. Critically, workflows such as IMPECCABLE combine large MPI simulations with large populations of smaller CPU- and GPU-bound tasks that must execute concurrently at campaign scale. Sustaining high utilization under these conditions requires the ability to place, overlap, and adaptively schedule heterogeneous workloads across very large resource pools, making heterogeneous workflows a foundational benchmark for RHAPSODY’s ability to scale efficiently on leadership-class HPC platforms.

\subsection{Inference at Scale}
\label{ssec:uc-inference}

Inference-at-scale workflows are characterized by performance being dominated by the execution of large numbers of inference requests on accelerator-resident models rather than by traditional simulation kernels. These workflows are increasingly common in scientific applications that employ large language models or deep neural architectures to guide analysis, generate hypotheses, or steer downstream computation. Their defining characteristics include high request concurrency, large input and output sequences, and the need to sustain high GPU utilization over extended periods.

Unlike conventional HPC workloads, inference-at-scale workflows are not expressed as a small number of long-running parallel jobs. Instead, they comprise large populations of independent inference requests issued in bursts or repeated rounds and serviced by multiple long-lived model instances. Performance is therefore governed less by individual inference latency than by the runtime system’s ability to host persistent inference services, distribute requests across them efficiently, and maintain sustained GPU occupancy without excessive CPU-side orchestration overhead.

The \emph{Low-Dose Understanding, Cellular Insights, and Molecular Discoveries (LUCID)} project exemplifies this class of workflows. In LUCID, long-context language models process large corpora of scientific literature to generate mechanistic hypotheses. Individual inference requests operate on thousands of tokens and incur substantial per-request GPU cost. To evaluate multiple models, prompting strategies, and configurations concurrently, inference requests are issued at high volume across multiple model instances, placing sustained pressure on GPU memory, scheduling, and request routing rather than on isolated model execution.

The \emph{SPHERICAL} (Scalable Platform for High-Throughput Inference in Computational Antigen design using Language models) campaign further illustrates inference at scale under realistic workflow conditions. Its execution combines a GPU-intensive inference workflow that submits large batches of independent requests to long-running model-serving services with a concurrent CPU-based scientific workflow that manages simulation and analysis tasks. Inference therefore proceeds under continuous background computation, requiring stable service behavior and predictable throughput under shared-resource conditions.

For RHAPSODY, inference-at-scale workloads stress the ability to support persistent accelerator-resident inference services, route large volumes of concurrent requests across multiple model instances, and sustain high GPU utilization over campaign durations while coexisting with other workflow components. As such, they provide a quantitative benchmark for RHAPSODY’s support of GPU concurrency, service-level parallelism, and throughput-driven execution on leadership-class HPC platforms.

\subsection{Coupled AI-HPC Workloads}
\label{ssec:uc-coupling}


Tightly coupled AI–HPC workflows form synchronous or semi-synchronous feedback loops in which simulations and AI components exchange data at high frequency. Unlike loosely coupled workloads, where inference or analysis can be executed asynchronously or offline, coupled workflows require low latency, strong data locality, and predictable end-to-end execution behavior. These requirements arise in workflows where AI steers simulations, evaluates intermediate states, guides sampling decisions, or replaces expensive subroutines with learned surrogates. In such settings, data movement and orchestration overhead often dominate time-to-solution.

Frameworks such as ROSE~\cite{rosealsaadi2025} and DeepDriveSim (DDSim) exemplify this class of workflows. ROSE orchestrates simulations in active-learning loops to construct and refine surrogate models, while DDSim enables AI-steered ensemble simulations in which learning models guide sampling and selectively trigger surrogate refinement during execution. Both require runtime support for low-latency coordination between simulation and AI components.

From a runtime perspective, coupled AI–HPC workflows pose several challenges. Data must be exchanged between simulation tasks and AI models with minimal overhead, synchronization boundaries must be managed efficiently, since simulations may block while waiting for inference results, and heterogeneous task lifetimes must be scheduled while preserving data dependencies. Because these interactions occur repeatedly within feedback loops, even modest orchestration inefficiencies are amplified across iterations.

The \emph{Integrated Machine-learning for PRotEin Structures at Scale} (IMPRESS) workflow illustrates these challenges in an agentic setting. IMPRESS employs an AI planning agent, implemented using Flowgentic~\cite{flowgentic} and LangGraph~\cite{wang2024agent}, that dynamically selects, parameterizes, and orders simulation, analysis, and learning tasks based on intermediate simulation outputs and accumulated metadata. This establishes a persistent AI-HPC control loop in which inference results influence the structure, scale, and resource profile of subsequent HPC tasks, requiring low-latency decision making and efficient execution of dynamically varying workloads.

Tightly coupled workflows also arise when AI training or inference is embedded directly within numerical simulations, such as in computational fluid dynamics~\cite{beck2021}. These cases are particularly sensitive to coupling latency due to the solvers' short per-iteration execution times. Inference introduces a two-way data exchange cost and can become a bottleneck in the simulation~\cite{Partee2022}, while differences in parallelization paradigms and resource requirements often necessitate separating AI and simulation components.

For RHAPSODY, tightly coupled workflows emphasize minimizing orchestration overhead while preserving data locality and supporting fine-grained synchronization. The runtime must coordinate simulation and AI tasks without introducing additional latency, while accommodating dynamically changing task graphs and resource demands driven by agentic decision making. As such, coupled workflows provide a benchmark for RHAPSODY’s ability to support synchronous and semi-synchronous AI–HPC interaction patterns at scale.

\section{Design and Implementation}\label{sec:design}

The design of RHAPSODY is motivated by a central observation: no single runtime system can efficiently support the diversity of execution models, performance characteristics, and heterogeneity required by hybrid AI–HPC workflows. Traditional HPC executables, high-throughput inference services, and tightly coupled AI–HPC computations differ fundamentally in their programming models, concurrency patterns, and data movement requirements, making monolithic runtime solutions either inefficient or overly restrictive when applied across all workload classes~\cite{brewer2024ai,jha2023ai}.

To address this gap, an ecosystem of specialized systems has emerged, each targeting a subset of these requirements. For example, Flux~\cite{ahn2014flux} and Dragon~\cite{DragonHPC, DragonHPCProxy} provide scalable mechanisms for launching and managing HPC executables and Python-based workloads; vLLM~\cite{kwon2023efficient} enables high-throughput inference serving; PyTorch~\cite{paszke2019pytorch} supports ML training and inference; and RADICAL-Pilot~\cite{merzky2021design} offers mature orchestration of heterogeneous tasks. While effective within their respective domains, none of these systems alone provides a unified execution substrate capable of concurrently supporting all workload classes at scale. RHAPSODY therefore adopts a building-blocks approach~\cite{turilli2019middleware}, integrating and extending existing runtimes rather than replacing them, and exposing their capabilities through uniform abstractions.
\jhanote{for some reason the spacing between paragraphs is greater than the previous session? could remove 5 lines if we can fix..}

Although RHAPSODY is inherently multi-runtime, this paper focuses on Dragon as the execution backend, building on prior demonstrations of concurrent Flux/Dragon integration at scale~\cite{merzky2025integrating}. Dragon’s architectural properties align closely with the \amnote{representative} type of AI–HPC workloads introduced in \S\ref{sec:uc}. This choice does not restrict RHAPSODY’s scope; rather, it enables us to concretely exercise all targeted workload classes while illustrating RHAPSODY’s abstractions for leadership-class systems. 

Dragon is a high-performance distributed runtime for Python-centric HPC workflows, providing scalable process management, distributed communication primitives, and shared-memory abstractions. Its API includes a fully compliant implementation of Python \texttt{multiprocessing} that scales to thousands of nodes, alongside constructs for distributed training, inference, telemetry, and workflow integration. Dragon supports traditional HPC environments (e.g., Slurm, PBS Pro) and cloud-like deployments via Kubernetes, and is optimized for Ethernet- and RDMA-enabled networks. Its support for highly dynamic processes and native HPC integration differentiates it from other distributed Python runtimes.

These properties make Dragon well-suited to heterogeneous workloads that combine large-scale simulations with fine-grained Python computations and inference services. Lightweight workers and distributed task queues allow RHAPSODY to co-schedule MPI jobs, GPU-accelerated kernels, and function tasks within a single allocation. Dragon’s shared- and distributed-memory execution backends further support scalable deployment of multi-instance inference services, enabling RHAPSODY to manage long-running AI services while sustaining high volumes of concurrent requests.

Finally, Dragon’s low-latency communication mechanisms align with the requirements of tightly coupled AI–HPC workloads. RHAPSODY builds on these capabilities to introduce coupling primitives that support direct data exchange between simulations and AI services. That minimizes redundant serialization and enables iterative, embedded interaction patterns. In this role, Dragon serves not as a privileged backend but as a representative runtime that allows us to demonstrate RHAPSODY’s general architectural approach.

\vspace*{-0.2em}
\subsection{Architecture}

\begin{figure}[!t]
    \centering
    \includegraphics[width=0.45\textwidth,height=7.3cm]{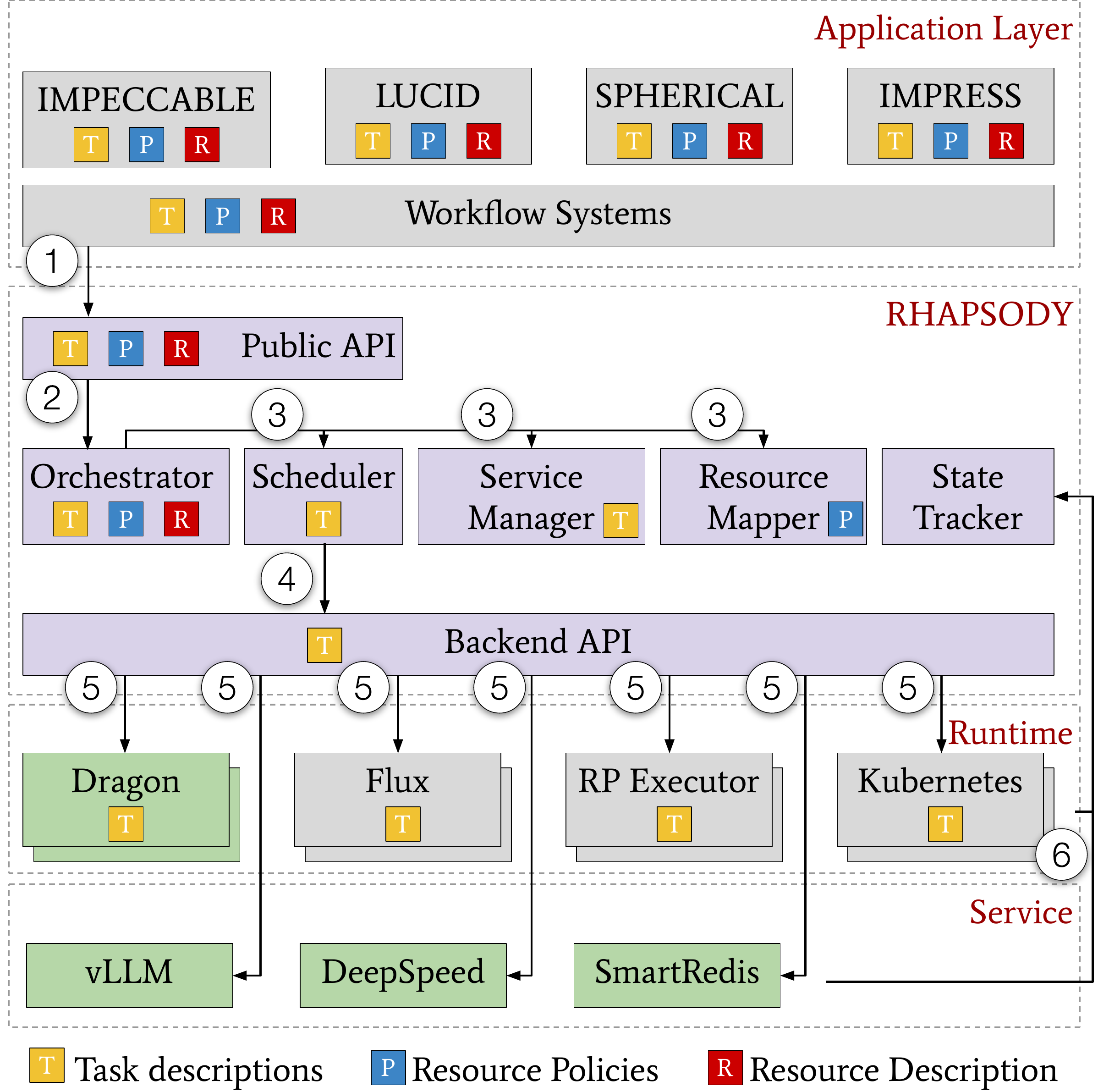}
    \caption{RHAPSODY architecture and execution model. Purple: RHAPSODY's components; Green: 3rd party systems used in this paper; Gray: 3rd party systems supported by RHAPSODY.
    \label{fig:rhapsody}
    \vspace*{-2.5em}}
\end{figure}

As shown in Fig.~\ref{fig:rhapsody}, RHAPSODY adopts a layered architecture that separates workflow-facing abstractions, middleware-level orchestration, and backend execution technologies. This enables clear separation of concerns and well-defined information flow across layers.

At the top layer, workflow systems and applications interact with RHAPSODY through a uniform, backend-agnostic public API. This API exposes three primary abstractions: task and resource descriptions, and execution policies. Task descriptions specify the computational payload, including execution semantics, dependencies, communication patterns, and per-task resource requirements, without exposing backend-specific details. Resource descriptions declaratively define the computational resources available to RHAPSODY for a given execution context, including CPUs, GPUs, and nodes. Execution policies express high-level constraints and preferences that guide how tasks and services are mapped onto available resources and backend runtimes.

This separation allows RHAPSODY to serve as a stable interface between diverse workflow systems---from static task graphs to adaptive, AI-driven workflows---and heterogeneous execution technologies. The task abstraction is intentionally general, supporting executables, language-level functions, long-running services, inference requests, and tightly coupled AI–HPC computations within a single execution model.

The middleware layer acts as the central orchestrator. It interprets task and resource descriptions under user- or system-defined policies, resolves dependencies, and determines execution strategies. The middleware maintains a global view of the execution environment, including available CPUs and GPUs, active services, data locality, and system load. This global state enables resource partitioning, concurrency control, backend selection, and service lifecycle management, supporting sustained concurrent execution within a single allocation.

Execution is delegated to backend runtime systems, which RHAPSODY enables to concurrently coexist within the same allocation. Backend runtimes provide concrete mechanisms for task and service execution, while RHAPSODY retains responsibility for orchestration, coordination, and policy enforcement. AI frameworks such as vLLM or DeepSpeed are treated as service workloads: they are instantiated through a backend runtime and subsequently managed by RHAPSODY’s middleware, which controls service lifecycle, readiness, endpoint discovery, and interactions with dependent tasks.

\vspace*{-0.5em}
\subsection{Execution Model}

RHAPSODY’s execution model generalizes the notion of a task to encompass four categories of computation common in hybrid AI–HPC workflows: (i) traditional HPC executables, including MPI simulations and GPU-accelerated kernels; (ii) language-level functions used for analysis or data reduction; (iii) long-running services, such as inference engines, memory stores, or agentic components; and (iv) tightly coupled AI–HPC tasks that require iterative data exchange between simulations and AI components. Interactions with persistent services are mediated through lightweight client tasks that communicate with service endpoints.

Fig.~\ref{fig:rhapsody} summarizes the execution flow. Workflow systems and applications submit task descriptions, resource descriptions, and execution policies through the public API (\circled{1}). These descriptions are ingested by the middleware, where orchestration, scheduling, service management, and resource mapping components interpret them using a global view of available resources and active services (\circled{2}--\circled{3}). Based on task type, resource requirements, and policy constraints, tasks and services are mapped to appropriate backend runtimes through the backend API (\circled{4}).

Backend runtimes execute tasks or instantiate services and report state transitions and events back to the middleware (\circled{5}). Dependent tasks discover and interact with services via middleware-managed endpoints, enabling inference calls and iterative AI–HPC coupling. Task and service lifecycles are tracked until completion or termination, allowing sustained concurrent execution and coordinated service management across the allocation (\circled{6}).

All task categories share a consistent middleware-level representation. RHAPSODY uniformly manages scheduling, dispatch, dependency resolution, and lifecycle transitions independently of the backend runtime used for execution. Services are treated as first-class entities that can be launched, monitored, and terminated explicitly, with their availability exposed to dependent tasks through middleware-managed discovery mechanisms. This uniform execution model allows workflow systems to reason about heterogeneous workloads without embedding backend- or framework-specific logic.

\vspace*{-0.5em}
\subsection{Resource Mapping and Concurrency Control}


Efficient utilization of heterogeneous resources is essential on HPC platforms. RHAPSODY provides a resource mapper that allocates CPUs, GPUs, and nodes to heterogeneous task types, including executables, language-level function tasks, and colocated services such as inference engines.

Resource mapping in RHAPSODY denotes the concrete assignment of task requirements---such as ranks, cores per rank, and GPUs per rank---to specific hardware resources. After mapping, each task rank is bound to a defined set of cores and accelerators. Mapping may be performed directly by RHAPSODY when the backend runtime supports explicit placement, or delegated to the backend when it provides its own intra-allocation scheduler. In this work, Dragon acts as the execution backend and supports resource mapping across the nodes it manages.

RHAPSODY can partition allocations into disjoint subsets of nodes and associate different execution backends with each partition. For example, one backend may manage partitions dedicated to multi-node MPI executables, while another manages partitions optimized for short-running function tasks or long-lived services. Tasks are classified by type and dispatched to the suitable backend, with scheduling decisions prioritizing balanced resource utilization over rigid concurrency limits.

Logical oversubscription is intentional. RHAPSODY maintains more ready tasks than can be executed concurrently so that newly freed resources can be backfilled immediately with compatible work. This backfilling-style execution model sustains high utilization despite heterogeneous task durations, resource footprints, and execution modalities. It directly targets hybrid AI–HPC workflows that combine large multi-node simulations, fine-grained analysis kernels, and colocated inference services within a single allocation, as described in \S\ref{sec:uc}.

\vspace*{-0.3em}
\subsection{Dragon Runtime Integration}




RHAPSODY integrates with Dragon to enable high-throughput execution of heterogeneous workloads on distributed workers. Dragon’s execution model supports batch execution of tasks with diverse resource requirements, including simulations, analysis kernels, and inference services. By executing both AI and HPC workloads atop the same scheduling and communication substrate, RHAPSODY ensures consistent task placement, data locality, and load balancing. 

\begin{figure}[!t]
    \centering
    \includegraphics[width=0.5\textwidth]{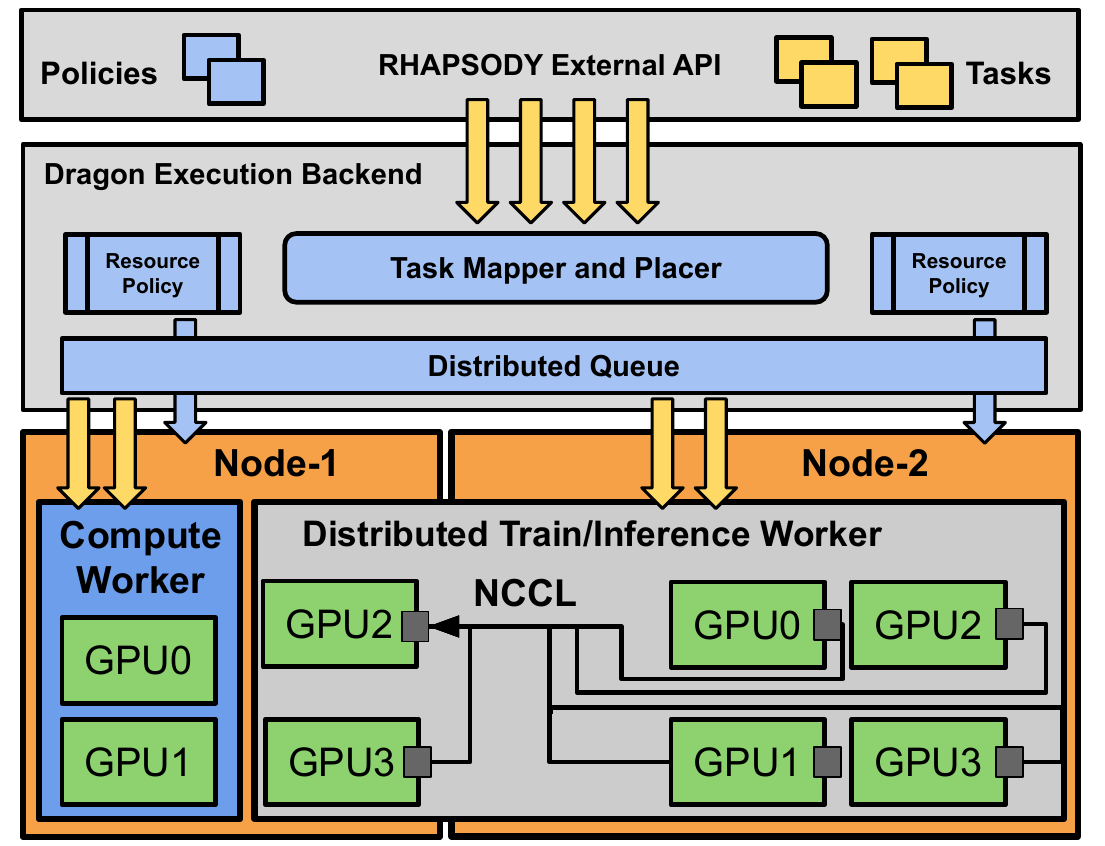}
    \caption{RHAPSODY–Dragon integration. Tasks and policies are submitted via the RHAPSODY external API and mapped by Dragon’s task mapper onto distributed queues and workers, enabling coordinated execution across nodes and GPUs.} 
    \label{fig:rhapsody_workers}
    \vspace*{-1.5em}
\end{figure}

To support distributed training and inference, RHAPSODY introduces specialized AI workers implemented as logical extensions of Dragon’s worker model (Fig.~\ref{fig:rhapsody_workers}). These workers coexist with compute workers and reuse Dragon’s native scheduling and communication services, while coordinating multi-GPU and multi-node execution for ML frameworks such as PyTorch and inference engines such as vLLM. This design enables on-demand, high-throughput AI capabilities to be embedded directly within HPC workflows.

RHAPSODY further extends Dragon’s execution backend through integration with SmartSim/SmartRedis~\cite{smartsim}, which provides an in-memory data exchange layer between simulations and AI components. Simulation state can be accessed by AI tasks as distributed in-memory objects, avoiding explicit data staging or duplication. This supports dynamic retraining and inference on live simulation data streams, enabling tightly coupled, iterative AI–HPC interaction patterns.

Together, these integrations allow RHAPSODY to compose Dragon’s distributed runtime, communication, and memory-management capabilities into an AI-aware middleware layer. Dragon remains responsible for executing tasks and services, while RHAPSODY orchestrates task composition, service coordination, and AI–HPC coupling across the workflow.

\pjmnote{Should this section say some specific things about what components from Dragon are being used? On one side, that may be too much technical detail, but without it are we being too vague about what "integration with Dragon" means?}\aanote{I rewrote the sub-section to be aligned with the title. I also tried to address Pete's comment}\jhanote{Aymen: I agree with Pete, some additional information about Dragon's internal would be good. I think Figure 3 sets this up nicely.}

\section{Performance Characterization \& Evaluation}\label{sec:exp}


Characterizing and evaluating RHAPSODY requires quantifying (i) system-level overheads introduced by orchestration and service management and (ii) the extent to which the runtime enables hybrid AI-HPC workflows to execute efficiently at scale. Because RHAPSODY targets concurrent execution of heterogeneous, possibly coupled AI-HPC workloads, our evaluation measures throughput, resource utilization, and scaling behavior on leadership-class platforms. Further, we consider heterogeneity width (HW), defined as the number of distinct task types executing concurrently, and the AI-HPC realization rate (ARR), defined as the rate at which AI-generated decisions are realized as executable HPC tasks.

We structure the evaluation around the scientific use cases introduced in \S\ref{sec:uc} and a set of focused system benchmarks. The experiments exercise complementary performance-critical capabilities: (1) scalable dispatch of large numbers of fine-grained tasks; (2) sustained concurrent execution of heterogeneous task types spanning different execution models, accelerator usage, and MPI scales; (3) high-throughput inference via long-running model-serving services; and (4) low-overhead, agent-driven coupling between AI components and simulations. Together, these experiments provide micro-level measurements (e.g., runtime overheads and response times) and macro-level outcomes (e.g., heterogeneity width (HW), AI-HPC realization rate (ARR), throughput, resource utilization, and strong and weak scaling).

We performed our experiments across multiple HPC platforms (OLCF Frontier, Purdue Anvil, and an HPE internal system) to demonstrate RHAPSODY's portability. The stack was deployed without platform-specific customization beyond enabling GPU support for ML backends.

\begin{table*}[t]
    \centering
    \caption{Experimental matrix. For each experiment: execution platform, workload type, runtime stack, task and node scale, and performance metrics. Abbreviations: A = AsyncFlow, D = Dragon, V = vLLM, R = ROSE, F = Flowgentic, RU = Resource Utilization, TH = Throughput, HW = Heterogeneity Width, OVH = Overhead, hom. = homogeneous, eth. = heterogeneous.}
    \label{tab:experiments}
    \footnotesize
    \begin{tabular}{
        p{6mm}@{\hspace{0mm}}
        p{24mm}@{\hspace{0mm}}
        p{18mm}@{\hspace{0mm}}
        p{31mm}@{\hspace{0mm}}
        p{15mm}@{\hspace{0mm}}
        p{43mm}@{\hspace{0mm}}
        p{16mm}@{\hspace{0mm}}
        p{23mm}
    } 
    \toprule
    \textbf{ID}                   &
    \textbf{Objective}            &
    \textbf{Platform}             &
    \textbf{Workload}             &
    \textbf{Stack}                &
    \textbf{\#Tasks}              &
    \textbf{\#Nodes}              &
    \textbf{Metrics}              \\
    \midrule
    1                             &    
    Scaling                       &    
    Anvil                         &    
    no-op                         &    
    A, D                          &    
    2,048--50,000 Functions       &    
    1--16                         &    
    RU, TH                        \\   
    2                             &    
    Heterogeneity                 &    
    Frontier                      &    
    Dummy sim.                    &    
    A, D                          &    
    295 Executable                &    
    256; 1,024                    &    
    HW                            \\   
    3                             &    
    Inference                     &    
    Anvil                         &    
    1--8,000 hom. prompts         &    
    A, D, V                       &    
    1--8 services; 10--80 clients &    
    1--8                          &    
    RU, TH                        \\   
    4                             &    
    Inference                     &    
    Perlmutter                    &    
    300 eth. prompts              &    
    A, D, V                       &    
    1--4 services                 &    
    1--4                          &    
    TH                            \\   
    5                             &    
    AI-HPC Coupling               &    
    HPE Internal                  &    
    Dummy sim. \& inf.            &    
    R, A, D, V                    &    
    100 sim. \& inf.; 1--512 Redis&    
    1--512                        &    
    Coupling data OVH             \\   
    6                             &    
    AI-HPC Coupling               &    
    Anvil                         &    
    Dummy Agents \& Tools         &    
    F, A, D, V                    &    
    4 services; 150--24,261 inf. \& tools &    
    4                             &    
    Coupling rate                 \\   
    \bottomrule
\end{tabular}
\end{table*}


\vspace*{-0.3em}
\subsection{Experiment 1: Baseline Runtime Performance}
\label{ssub:exp-overheads}


In this experiment, we characterize RHAPSODY’s baseline runtime behavior using null workloads, independently of any application-specific logic. The goal is to quantify system-level overheads associated with task dispatch, scheduling, and coordination, which are exercised to different degrees by the scientific use cases described in \S\ref{sec:uc}. We evaluate the strong- and weak-scaling behavior of RHAPSODY integrated with the Dragon runtime on Purdue Anvil compute nodes.

For weak scaling, the total number of independent, non-operational (no-op) CPU-based function tasks increases proportionally with the number of nodes and RHAPSODY-managed workers, ranging from 2,048 to 32,768 tasks, thereby maintaining a constant workload per node. For strong scaling, the total workload is fixed at 50,000 no-op tasks while the number of compute nodes increases from 1 to 16. We report two metrics: throughput (tasks/s), defined as the number of completed tasks per unit time, and runtime overhead (seconds), defined as the cumulative time spent by RHAPSODY and Dragon to submit, schedule, execute, and collect all tasks.

\begin{figure}[!t]
    \centering
    \includegraphics[width=0.5\textwidth]{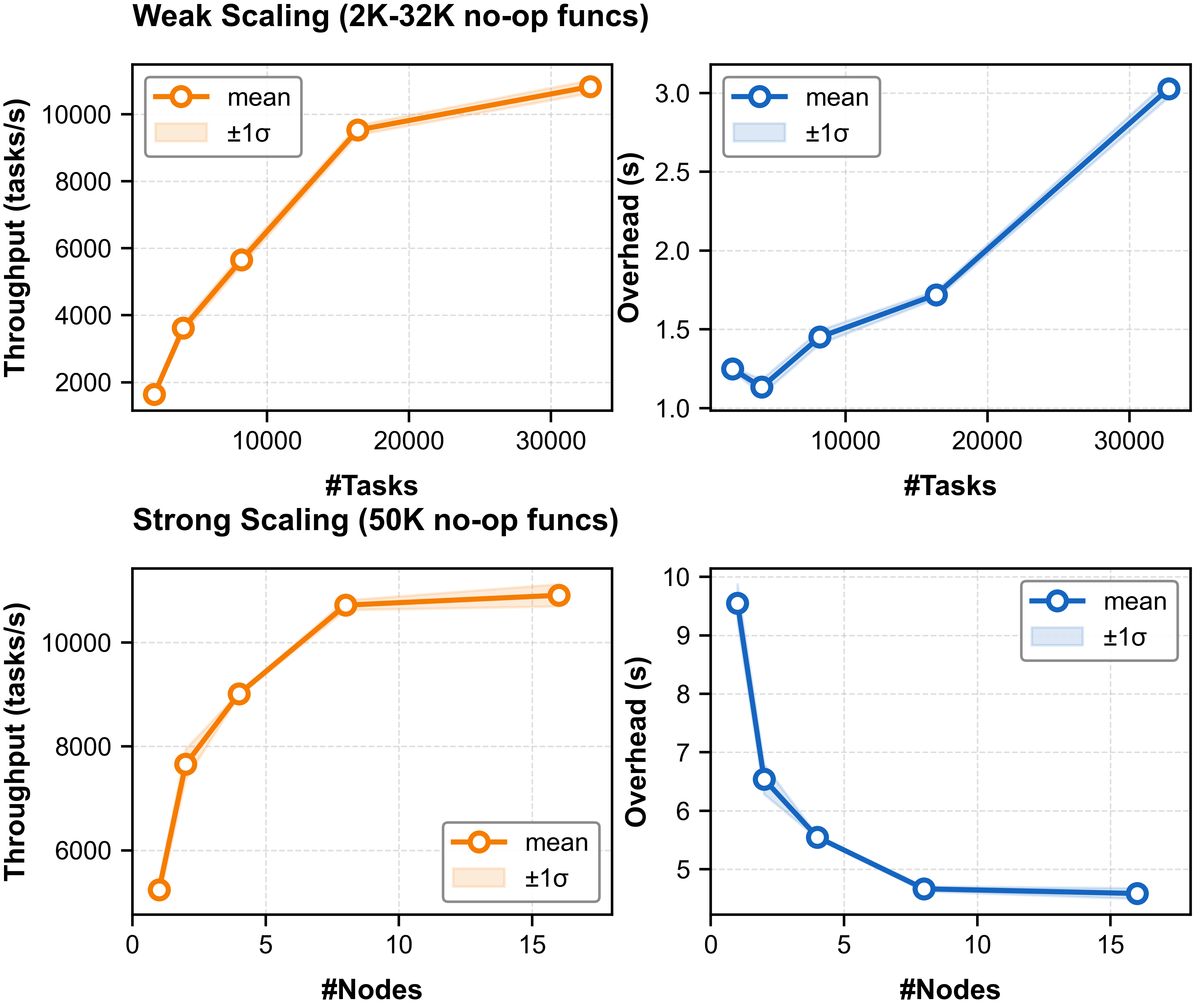}
    \caption{RHAPSODY scaling performance with Dragon runtime system when executing $N$ single-core no-op function tasks.}
    \label{fig:rhapsody_scale}
\end{figure}

Fig.~\ref{fig:rhapsody_scale} shows average throughput and runtime overhead across 1--16 nodes. Under weak scaling (top), RHAPSODY achieves near-linear efficiency. Throughput increases proportionally to the combined growth in workload and resources, exceeding 10,000~tasks/s. Runtime overhead remains low ($\sim$1--3s) and grows only marginally as task counts increase, indicating that coordination costs do not dominate execution even for large batches of fine-grained tasks (corresponding to approximately 100--300~$\mu$s of overhead per task).
\pjmnote{it might be helpful to the reader to state a no-op task is completed in about 100-300 microseconds}\mtnote{Better?}

Under strong scaling (bottom), throughput increases from $\sim$5,000 to 11,000~tasks/s as node counts grow from 1 to 8, while aggregate runtime overhead decreases by roughly 50\%, reflecting reduced queuing and submission delays at higher degrees of parallelism. Beyond 8 nodes, throughput plateaus near 11,000~tasks/s, indicating that peak parallel efficiency has been reached for this workload size.

Overall, this experiment demonstrates that RHAPSODY introduces minimal overhead when managing large numbers of short-lived tasks and scales efficiently under both strong and weak scaling regimes. These properties are essential for hybrid AI-HPC workflows that decompose computation into thousands of fine-grained operations, such as distributed inference, data-parallel training, and agent-driven task generation. Further, these results provide a baseline for interpreting the application-level results in the following experiments.

\subsection{Experiment 2: Heterogeneous Workflows}
\label{ssub:exp-heterogeneity}


This experiment characterizes and evaluates RHAPSODY’s ability to sustain the concurrent execution of heterogeneous workloads comprising tasks with different execution models, accelerator usage, and MPI scales. The objective is to determine whether RHAPSODY can coordinate many distinct task types within a single campaign without imposing artificial execution phases or serializing heterogeneous work.

We use a representative campaign derived from the heterogeneous use case described in \S\ref{ssec:uc-heterogeneity}, in which multiple workflows execute concurrently within a shared allocation. Tasks have three dimensions of heterogeneity: (1) execution model (serial vs.\ MPI); (2) accelerator usage (CPU vs.\ GPU); and (3) MPI scale. Each unique combination of these properties defines a distinct task type. This definition yields an application-agnostic, quantitative measure of workload heterogeneity.

Across the campaign, task types include serial CPU and GPU tasks, CPU-based MPI tasks, and GPU-based MPI tasks spanning a wide range of scales. MPI tasks range from 32 to 7,168 ranks and typically execute significantly longer than serial tasks, resulting in a large variation in resource footprints and task durations. Task submission and execution order are determined solely by workflow dependencies rather than by runtime-imposed scheduling phases. All experiment runs were performed on OLCF Frontier using 256 and 1,024 nodes to assess whether RHAPSODY’s handling of heterogeneity remains stable as scale increases. In both configurations, multiple workflows execute concurrently within the same allocation.

Fig.~\ref{fig:impeccable} shows the number of distinct task types executing concurrently over time, which we denote as \emph{heterogeneity width (HW)}, defined as the number of distinct task types executing concurrently at a given time. HW is bounded by task dependencies and resource feasibility, and therefore provides a relative measure of how effectively the runtime exploits available heterogeneity rather than an absolute optimum. At 256 nodes, HW supports up to 11 concurrent task types, with most executions sustaining between 1 and 4 types. At 1,024 nodes, the peak HW increases to $\sim$22 task types, with extended intervals sustaining more than 6 concurrent types. 

These results indicate that larger allocations enable broader and persistent overlap among heterogeneous execution modes. A poor outcome would show persistently low HW independent of scale, indicating artificial phase separation or serialization imposed by the runtime rather than by workflow dependencies.

\iftrue
\begin{figure}[!t]
    \centering
    \includegraphics[width=\linewidth]{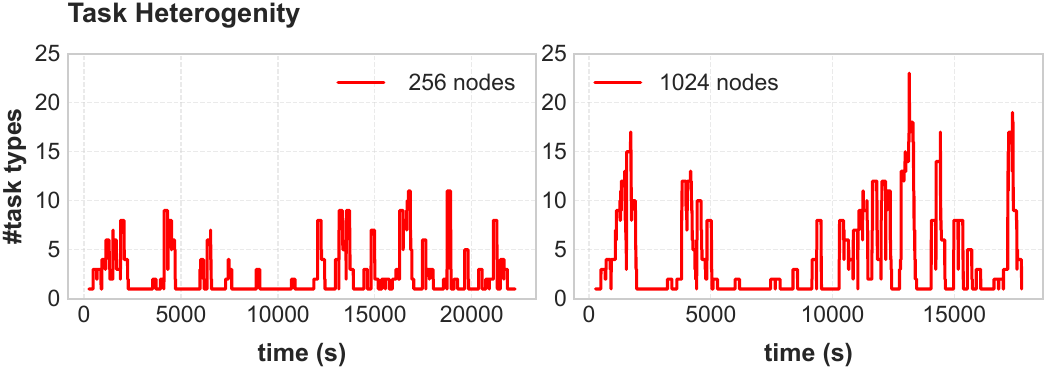}
    \vspace*{-1.5em}
    \caption{
    Runtime dynamics of task heterogeneity under RHAPSODY. Instantaneous heterogeneity width (HW), defined as the number of distinct task types executing concurrently, over time for a heterogeneous campaign executed on (top) 256 nodes and (bottom) 1,024 nodes. Higher HW indicates greater overlap among heterogeneous execution modes.}
    \label{fig:impeccable}
    \vspace*{-1.5em}
\end{figure}
\fi

In both configurations, RHAPSODY does not cluster execution into homogeneous phases. Long-running MPI tasks form a persistent execution background, while shorter serial CPU and GPU tasks are launched opportunistically as dependencies are satisfied. Periods of reduced HW reflect the structure and readiness of the workload rather than limitations imposed by the runtime. Importantly, increasing scale does not reduce heterogeneity or introduce phase separation; instead, it expands the range and duration of concurrent task-type execution, according to the greater amount of resources available.

Overall, this experiment demonstrates that RHAPSODY can coordinate campaigns comprising many heterogeneous task types within a single allocation, including tasks spanning a range of MPI scales. By preserving workflow semantics and avoiding runtime-imposed serialization, RHAPSODY enables heterogeneous width to grow naturally with available resources, a key requirement for hybrid AI-HPC workflows.

\subsection{Experiments 3--4: Inference at Scale}
\label{ssub:exp-inference-scale}

This section characterizes and evaluates RHAPSODY’s ability to support high-throughput inference at scale, following the use cases described in \S\ref{ssec:uc-inference}. We focus on inference workloads executed via long-running vLLM services coordinated by RHAPSODY and Dragon, and on the runtime mechanisms required to sustain high concurrency, efficient request routing, and scalable performance under heterogeneous inference costs.


We performed Experiment 3 on Purdue Anvil and Experiment 4 on NERSC Perlmutter. Each configuration consists of a set of compute nodes hosting vLLM service instances and a set of inference clients that asynchronously submit batched requests. Clients and services are not co-located, and requests are distributed to service instances via network calls using runtime-managed routing. Inference clients execute as CPU-side tasks, while model execution occurs on GPUs within service instances. Services and clients are managed by RHAPSODY as illustrated in Figs.~\ref{fig:rhapsody} and~\ref{fig:rhapsody_workers}.\aanote{addressed Pete's last comment about the co-location of inference clients and services}

We first evaluate baseline scalability by proportionally increasing the number of nodes, service instances, and inference clients. Specifically, we consider configurations of 1/1/10, 2/2/20, 4/4/40, and 8/8/80 (nodes/service instances/inference clients), preserving a constant ratio between request sources and model-serving capacity. This isolates runtime scalability effects from workload imbalance.


\begin{figure*}[!t]
    \centering
    \begin{subfigure}[t]{0.24\linewidth}
        \centering
        \includegraphics[width=\linewidth]{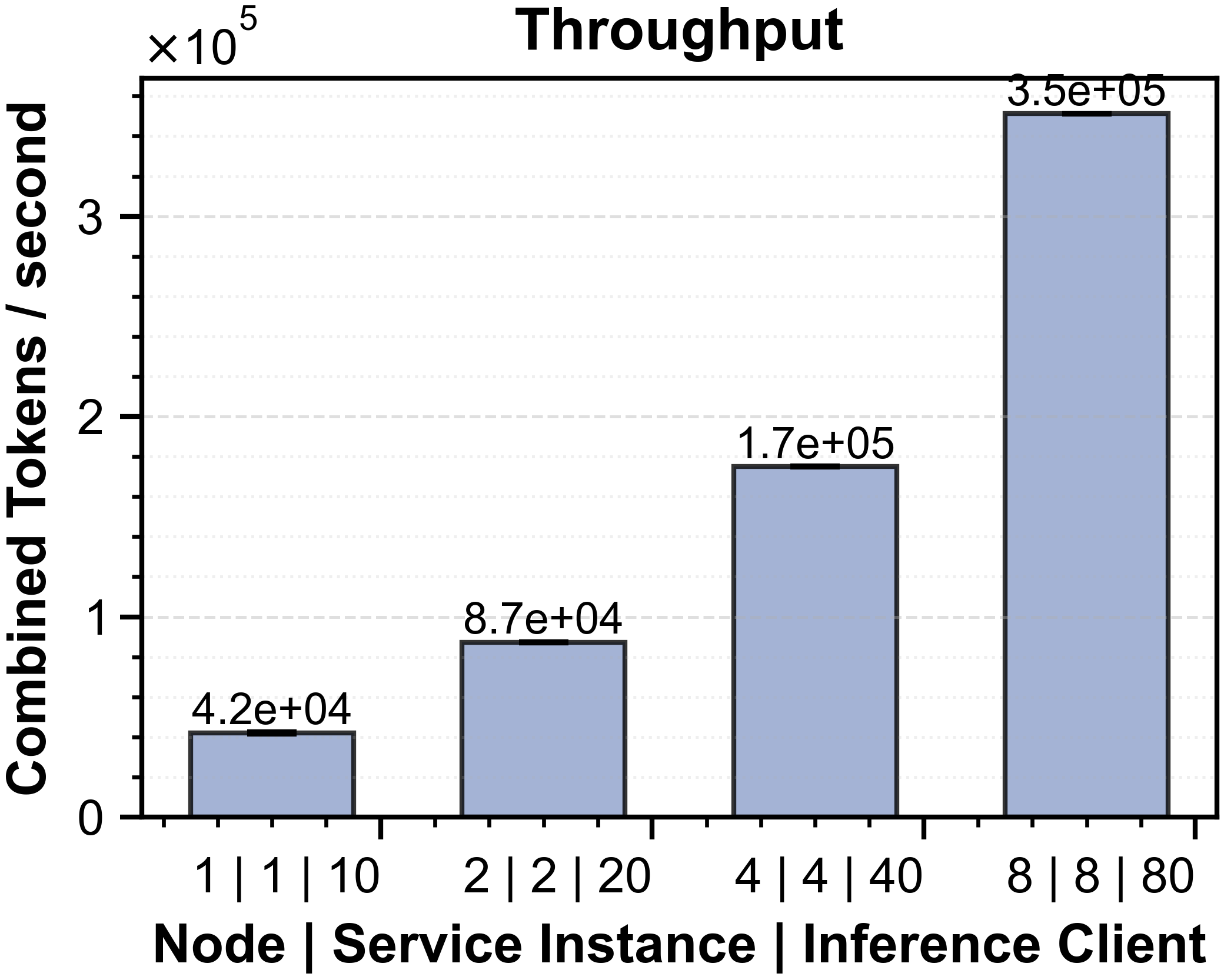}
        \caption{}
        \label{fig:spherical_a}
    \end{subfigure}\hfill
    \begin{subfigure}[t]{0.24\linewidth}
        \centering
        \includegraphics[width=\linewidth]{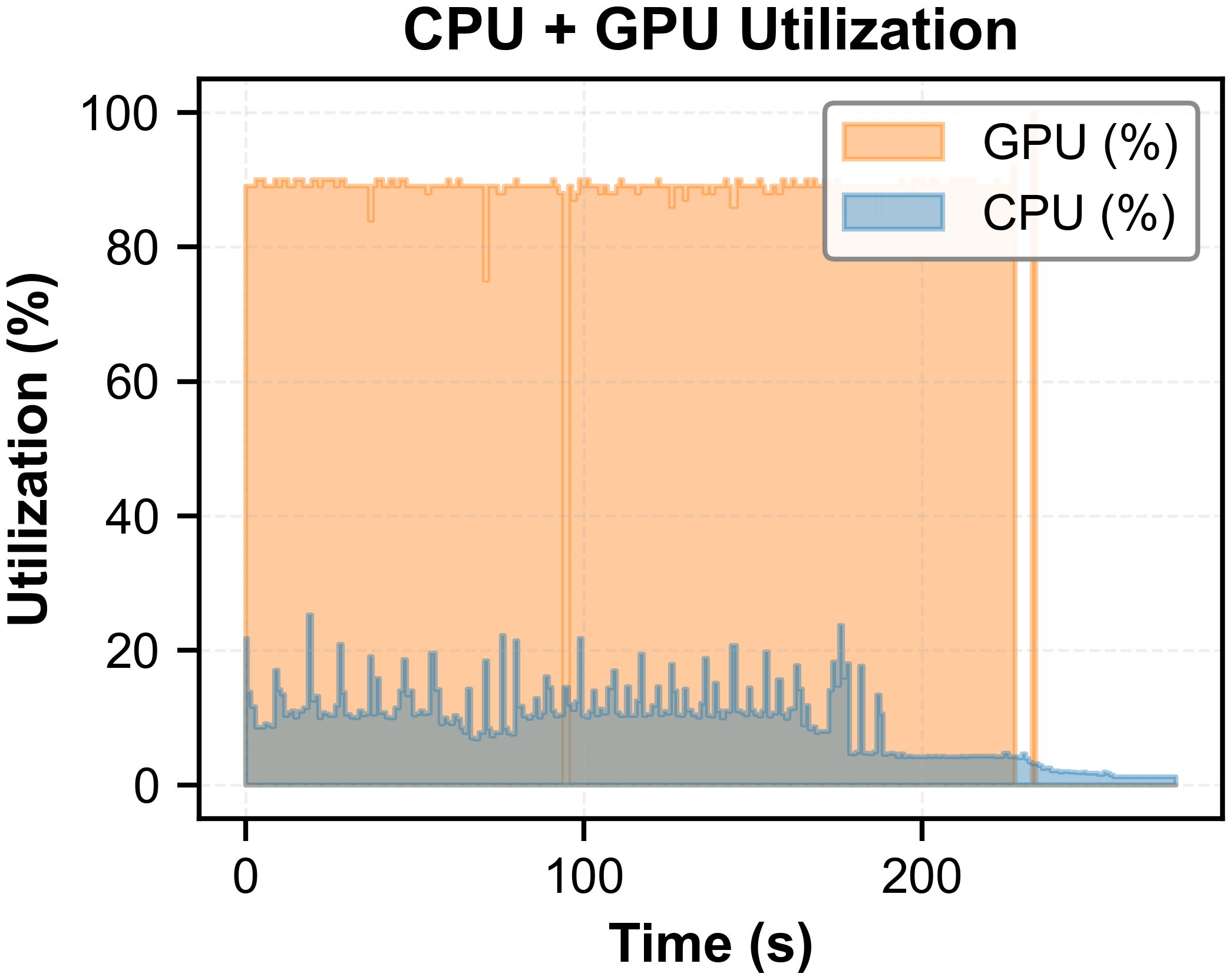}
        \caption{}
        \label{fig:spherical_b}
    \end{subfigure}\hfill
    \begin{subfigure}[t]{0.24\linewidth}
        \centering
        \includegraphics[width=\linewidth]{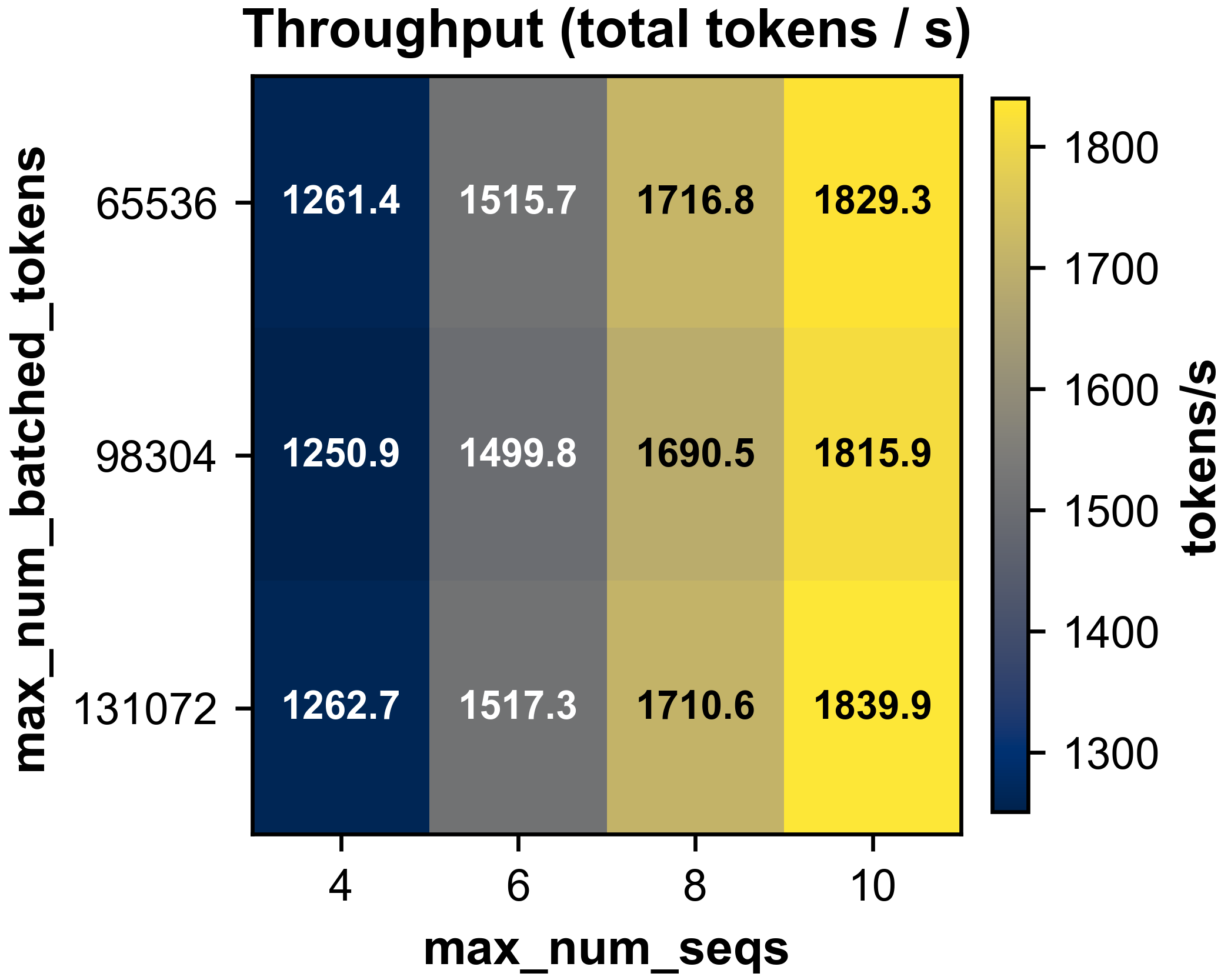}
        \caption{}
        \label{fig:lucid_c}
    \end{subfigure}
    \begin{subfigure}[t]{0.24\linewidth}
        \centering
        \includegraphics[width=\linewidth]{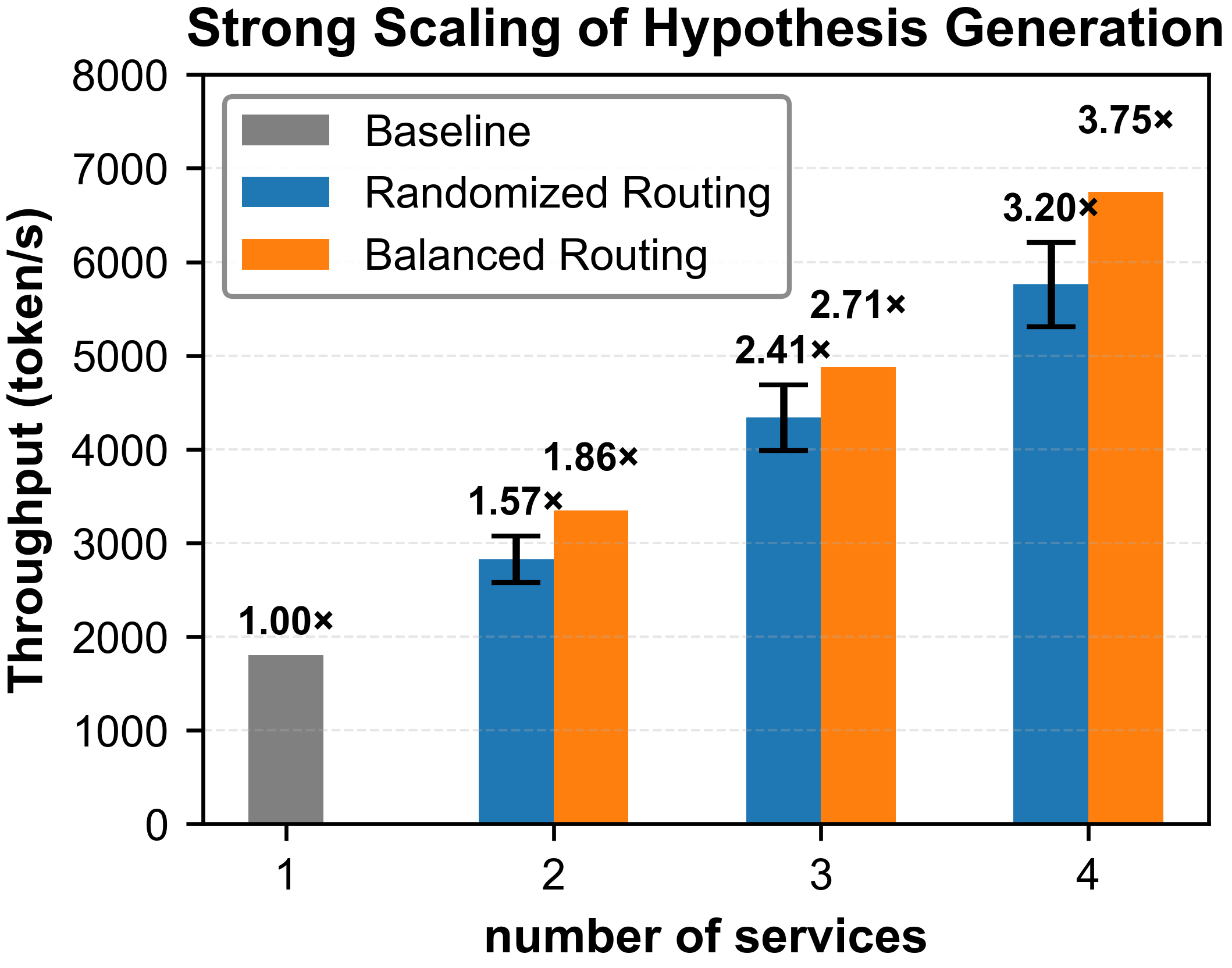}
        \caption{}
        \label{fig:lucid_d}
    \end{subfigure}
    \caption{Scalability of high-throughput inference under RHAPSODY (a, b) and configuration sensitivity and strong scaling of multi-service inference under heterogeneous request costs (c, d).
    (a) Aggregate token throughput as the number of service instances and inference clients increases proportionally.
    (b) Representative CPU/GPU utilization trace for one node, illustrating sustained GPU utilization during asynchronous inference execution.
    (c) Sensitivity of total inference throughput to vLLM batching parameters.
    (d) Strong scaling of aggregate inference throughput with randomized and token-aware routing policies.
    }
    \label{fig:spherical_combined}
\end{figure*}

Fig.~\ref{fig:spherical_combined} (a,b) shows aggregate inference throughput and resource utilization. GPU utilization is consistently high ($\sim$90--95\%) across all configurations, indicating that RHAPSODY sustains a steady flow of inference requests and avoids GPU starvation as concurrency increases. Aggregate throughput scales from $4.2\times10^{4}$ to $3.5\times10^{5}$~tokens/s across one to eight nodes and exhibits near-linear scaling with the number of service instances. This confirms effective utilization of additional GPU capacity without orchestration-induced bottlenecks.

We next evaluate inference under heterogeneous request costs using a hypothesis-generation workload with extreme variability in prompt length ($\sim$4,000--50,000 tokens). Such variability makes naive request distribution prone to load imbalance and straggler effects. To identify efficient operating regimes, we first explore vLLM batching parameters using a fixed subset of prompts. Fig.~\ref{fig:spherical_combined} (c) shows that throughput is primarily influenced by \texttt{max\_num\_seqs}, while remaining comparatively insensitive to \texttt{max\_num\_batched\_tokens}. This demonstrates RHAPSODY’s ability to orchestrate parallel exploration of inference configurations when workflows are sufficiently large to amortize setup costs.


Finally, we evaluate strong scaling by replicating inference services and distributing a fixed set of 300 prompts across instances. Aggregate throughput is computed as the total number of processed tokens divided by the end-to-end makespan. We compare randomized routing, which assigns prompts uniformly at random, with token-aware balanced routing, which equalizes both prompt count and estimated input-token volume per service. Fig.~\ref{fig:spherical_combined} (d) shows that balanced routing consistently achieves higher scaling efficiency and lower variance by reducing straggler effects. These results indicate that, for irregular long-context inference workloads, scalable throughput depends on both service-level parallelism and runtime-level routing policies.

Overall, these experiments demonstrate that RHAPSODY effectively supports inference-at-scale workloads by combining persistent inference services with scalable orchestration and flexible request routing. By sustaining high GPU utilization and near-linear throughput scaling under both uniform and highly irregular request costs, RHAPSODY enables efficient large-scale inference on HPC platforms.

\subsection{Experiments 5--6: Coupled AI-HPC Workloads}
\label{ssub:exp-hpc-ai-coupling}


These experiments characterize and evaluate RHAPSODY’s support for tightly coupled AI-HPC workloads, focusing on data-movement overheads and on the temporal coupling between AI-driven decisions and HPC task execution. The experiments correspond to the coupled use cases described in \S\ref{ssec:uc-coupling} and exercise both data-plane and control-plane interactions under increasing scale.

We first analyze coupling overheads in a representative simulation-inference workflow that exchanges data at runtime. Each configuration executes 100 simulation-inference pairs per node. Simulations produce 4,000-element tensors ($\sim$16~KB per task, 1.6~MB per node), and inference tasks consume this data. To isolate data-movement and orchestration costs, both simulations and inference tasks are implemented as lightweight stand-ins whose runtime is dominated by data transfer rather than computation.

We compare two coupling mechanisms: filesystem-based exchange via a RAM disk and memory-based exchange using Redis deployed through SmartRedis. For the Redis-based configuration, RHAPSODY launches one Redis instance per node and pins simulation, inference, and datastore processes to preserve locality. Experiments run on 1--512 nodes.

\begin{figure*}[t]
    \centering
    \includegraphics[width=1\textwidth, height=4.2cm]{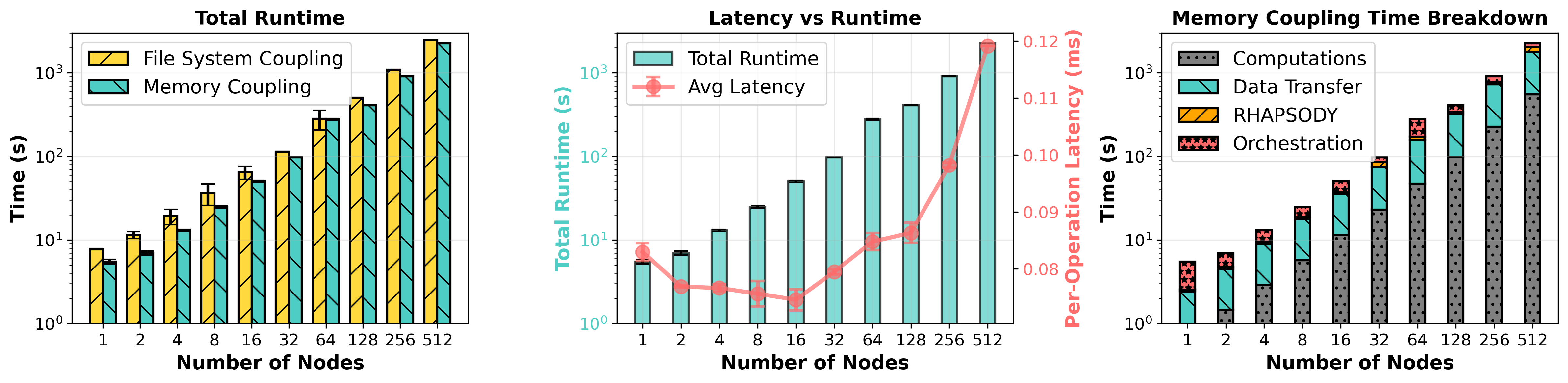}
    \caption{Coupling and data transfer overheads in RHAPSODY-enabled coupled AI-HPC workflows from a strong-scaling study (data volume was 1.6MB per node). The total data volume Left: Total runtime for memory- vs. filesystem-based coupling across 1--512 nodes. Middle: Relationship between average \texttt{PUT}/\texttt{GET} latency and end-to-end workflow runtime. Right: Memory-coupled runtime decomposition into computation, data transfer, Orchestration, and RHAPSODY overhead.}
    \label{fig:rhapsody_couple_overheads}
\end{figure*}

Figure~\ref{fig:rhapsody_couple_overheads} (left) shows total workflow runtime for both coupling mechanisms. Memory-based coupling consistently outperforms filesystem-based exchange at all scales, with up to a 50\% reduction in runtime at 8 nodes. Scaling trends are smooth across node counts, indicating that RHAPSODY does not introduce disruptive scheduling or queuing effects.

The middle panel reports average \texttt{PUT}/\texttt{GET} latency measured via SmartRedis. Latency increases only slightly, from $\sim$0.08 to 0.12~ms across 1--512 nodes, despite a fourfold increase in the number of datastore operations. Because data movement is handled directly by SmartRedis, these measurements reflect datastore behavior rather than RHAPSODY overhead and demonstrate that RHAPSODY does not amplify communication latency.

The right panel decomposes the memory-coupled runtime into computation, data transfer, workflow orchestration, and RHAPSODY overhead. As scale increases, data transfer becomes the dominant cost, growing proportionally with total data volume (0.32~GB at 1 node; 164~GB at 512 nodes). In contrast, orchestration and RHAPSODY overheads remain small, contributing $<$5\% of total runtime across all scales.

We next evaluate AI-HPC coupling behavior using a representative workflow that emulates AI-driven decision making followed by HPC task execution, based on the agentic use case in \S\ref{ssec:uc-coupling}. Using execution traces, we compute two windowed rates over time: (i) the \emph{agent decision rate}, defined as the number of LLM inference requests issued per unit time; and (ii) the \emph{AI-HPC realization rate (ARR)}, defined as the number of HPC tasks transitioning into the \texttt{RUNNING} state per unit time. Rates are computed over a sliding time window to expose coupling dynamics while smoothing per-event noise.

\begin{figure*}[t]
    \centering
    \includegraphics[width=0.45\textwidth,height=4.5cm]{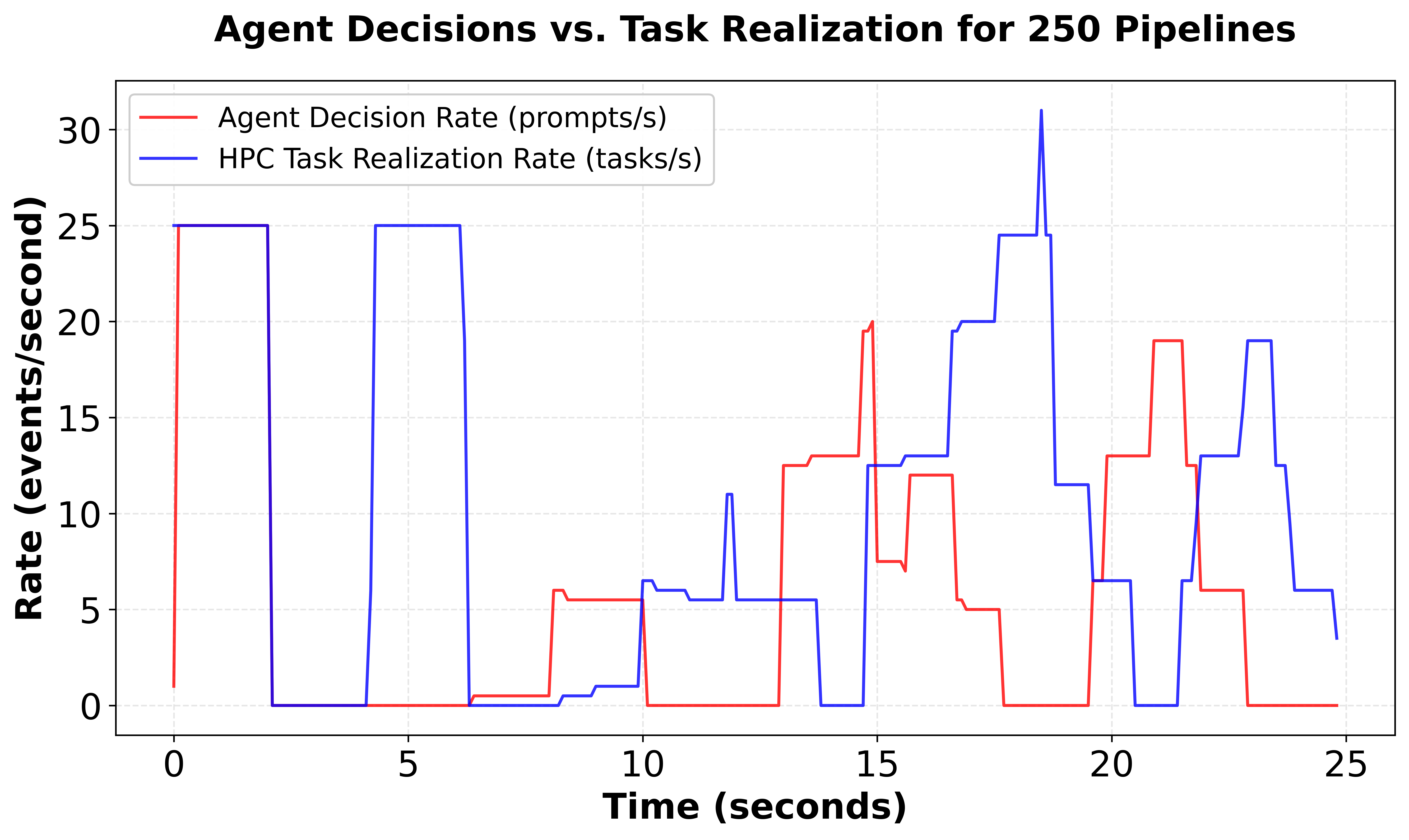}
    \includegraphics[width=0.45\textwidth,height=4.5cm]{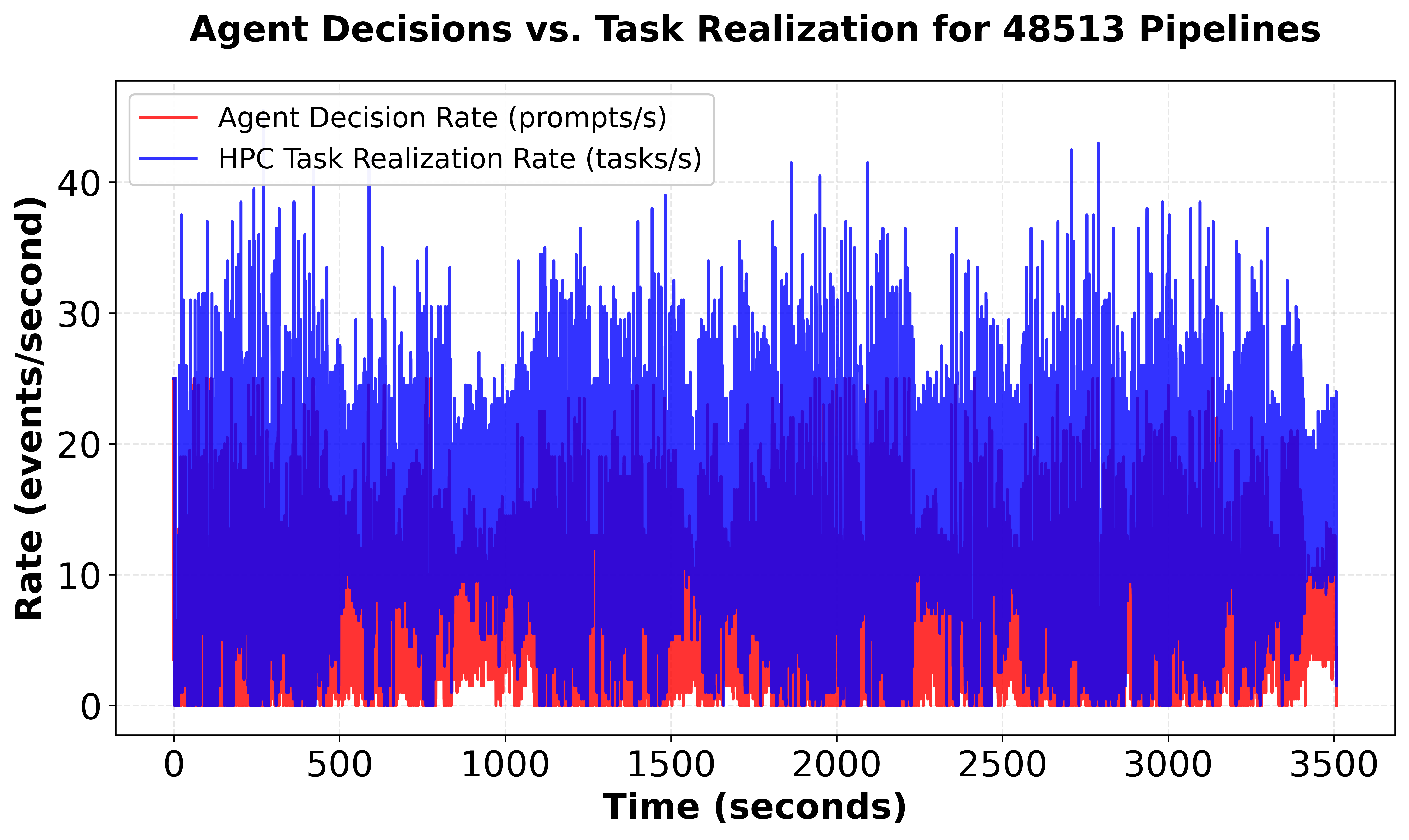}
    \caption{Agent decision rate vs. HPC task realization rate in an agentic AI–HPC workflow, measured over 25s with 250 agents (left) and 3,500s (right) with~$\sim$49,000 agents. Events = agent decisions and HPC task realization. Sustained overlap and bounded lag indicate stable and effective AI–HPC coupling under RHAPSODY.}
    \label{fig:coupling_rates}
    \vspace*{-1.5em}
\end{figure*}

Figure~\ref{fig:coupling_rates} shows these rates for short and long executions under increasing agent population. In both cases, increases in agent decision rate are followed by corresponding increases in ARR with bounded lag. This indicates that AI-generated decisions are promptly realized as executable HPC work. The two rates show sustained temporal overlap, rather than phased execution, showing that RHAPSODY does not impose artificial serialization between AI reasoning and HPC execution.

As execution proceeds, neither rate collapses under load: even during bursty decision phases, ARR remains active and stable. Moreover, the traces reveal emergent feedback behavior characteristic of agentic control loops, in which periods of high task realization are associated with moderated decision rates. This feedback arises from workflow logic rather than from runtime-imposed constraints, indicating that RHAPSODY preserves agent-level control semantics.

Overall, these experiments show that RHAPSODY enables efficient execution of tightly coupled AI-HPC workflows by preserving data locality, minimizing orchestration overhead, and ensuring that control-plane activity is consistently and efficiently realized in the execution plane. By maintaining bounded lag and sustained concurrency between AI decisions and HPC task execution, RHAPSODY supports synchronous and semi-synchronous AI-HPC interaction patterns at scale.

\section{Related Work}\label{sec:related}

\mtnote{I propose an iteration that slightly clarified the boundaries between workflows and runtime systems, adding a few references but respecting the narrative already present.}

Several frameworks and systems have been developed to support the integration of AI with traditional HPC workloads. Most existing efforts, however, focus on specific coupling patterns, application domains, or execution models, rather than providing a general-purpose runtime substrate for hybrid AI-HPC workflows that combine simulation, training, inference, and agent-driven control at scale.

smiBench targets the deployment and benchmarking of machine-learned surrogate models on HPC platforms using inference servers~\cite{brewer2021production}. While it provides useful primitives for simulation-surrogate workflows, it does not address the runtime requirements of dynamic, multi-stage AI-HPC workflows that integrate heterogeneous tasks, services, and adaptive execution logic within a single allocation.

FIRST enables inference-as-a-service for large language models on HPC systems by integrating multiple LLM serving tools behind a unified API to support high-throughput inference~\cite{10.1145/3731599.3767346}. Although FIRST addresses scalable inference delivery, it focuses on inference as an isolated capability. In contrast to general-purpose serving systems such as Triton Inference Server~\cite{nvidia_triton} or Ray Serve~\cite{ray_serve}, FIRST targets HPC-native deployment but does not address broader coordination of heterogeneous AI-HPC workloads or their interactions with simulation-driven execution patterns. RHAPSODY complements this space by supporting inference services as first-class runtime entities that coexist with HPC executables and adaptive workflows.

The eFlows4HPC project introduces workflow abstractions and a software stack for composing HPC simulations, data analytics, and machine learning~\cite{Ejarque_2022}. While it supports dynamic workflow construction, it does not integrate high-throughput inference services into the runtime layer nor address service-oriented execution and coordination within a single allocation. RHAPSODY instead targets the runtime level, enabling concurrent execution of heterogeneous workloads and persistent services across multiple backend technologies.

Workflow systems such as Parsl~\cite{babuji2019parsl} or Pegasus~\cite{deelman2015pegasus} provide high-level abstractions for composing large-scale scientific workflows on HPC systems. Several application- and workflow-level frameworks focus on specific coupled AI-HPC patterns. DRLFluent couples distributed reinforcement learning with a CFD solver to support in-situ learning and control~\cite{MAO2023102171}, while Colmena enables ML-driven steering of ensemble simulations using Parsl for task execution and retraining~\cite{Ward_2021}. Similarly, systems such as ROSE and DeepDriveSim support active-learning and AI-steered simulation campaigns. These efforts build on earlier work in in-situ learning and steering, where ML components are embedded into simulation workflows to reduce data movement and latency~\cite{maulik2021}.

Recent work has also explored agentic execution models, in which LLM-based agents dynamically generate and coordinate computational tasks based on intermediate results and external state~\cite{wang2024agent,flowgentic}. These systems operate primarily at the workflow or application level and are often specialized to particular coupling patterns or control structures.

In contrast, RHAPSODY targets the runtime layer rather than a specific workflow model or application pattern. It provides general abstractions for tasks, services, resources, and coupling that workflow systems, agentic frameworks, and domain-specific applications can leverage without embedding runtime-specific logic. In this sense, RHAPSODY complements prior work on pilot-based and multi-level scheduling systems~\cite{merzky2021design,ahn2014flux} by extending runtime support to service-oriented execution and agent-driven AI-HPC interaction.

\jhanote{line spacing between paragraphs is large here again}\mtnote{Fixed now, it is usually an issue with using too many vspaces.} 

Overall, RHAPSODY is not intended to replace existing workflow frameworks, inference servers, or backend runtimes. Instead, it provides a unifying runtime substrate that enables these components to interoperate and execute concurrently at scale, supporting heterogeneous, service-oriented, and tightly coupled AI-HPC workflows on leadership-class systems.

\section{Conclusions}\label{sec:conclusion}

Hybrid AI-HPC workflows increasingly combine large-scale simulation, high-throughput inference, and tightly coupled, agent-driven control within a single execution campaign. Supporting these workflows requires runtime systems that can coordinate heterogeneous execution models, persistent services, and fine-grained coupling without imposing artificial phases or incurring prohibitive overheads.

This paper introduced RHAPSODY, a multi-runtime middleware that integrates existing HPC runtime and AI serving systems through uniform APIs and abstractions for tasks, services, and resources. By composing rather than replacing specialized runtimes, RHAPSODY enables concurrent execution of heterogeneous workloads, scalable inference services, and synchronous AI-HPC interaction within a single allocation.

Our evaluation on multiple HPC platforms shows that RHAPSODY: (1) introduces minimal runtime overhead; (2) sustains high workload heterogeneity at scale; (3) achieves near-linear scaling for inference-based workloads; (4) enables efficient data coupling between AI and HPC tasks; (5) preserves low-latency coupling between AI decision making and HPC task execution; and (6) in agentic workflows, ensures that control activity is consistently realized in the execution plane, maintaining bounded lag and stable concurrency.

Our results show that RHAPSODY provides a practical and scalable foundation for executing emerging hybrid AI-HPC workflows on leadership-class HPC platforms. This enables heterogeneous, service-oriented, and tightly coupled execution motifs to coexist, maintaining performance and scalability. RHAPSODY will soon be available and used on the American Science Cloud to support hybrid AI-HPC workflows.


 \section*{Acknowledgments}
\noindent\footnotesize{{{\bf Experiments} Data and analysis at: {\footnotesize \url{https://github.com/radical-experiments/rhapsody}}}}

\small
\bibliographystyle{IEEEtran}
\bibliography{rhapsody}

\end{document}